\documentclass{article}
\usepackage{jim,amsmath}
\usepackage{amssymb}
\usepackage{amsfonts}
\usepackage[utf8]{inputenc}
\usepackage[T1]{fontenc}
\usepackage{graphicx}

\usepackage{url}
\usepackage{hyperref}
\usepackage{listings}
\usepackage{courier}

\usepackage{ifthen}

\usepackage{caption}
\usepackage{subcaption}

\usepackage{svg}
\usepackage{pgfplots}
\usepackage{tikz}
\usetikzlibrary{math} 
\usetikzlibrary{patterns}
\usetikzlibrary{decorations.pathmorphing}
\usetikzlibrary{arrows,arrows.meta,automata,shadows,shapes,shapes.geometric,circuits.logic.US,shapes.gates.logic.US,calc}
\usetikzlibrary{positioning,backgrounds,fit}

\newcommand{\tkzscalex}{4.5ex}
\newcommand{\tkzscaley}{6ex}

\tikzset{
  >=Stealth, 
  multrectangle/.style={draw, fill=black!5},
  generator/.style={draw,align=center,fill=blue!5,rounded corners, minimum width=15ex, minimum height=8ex},
  hwblock/.style={draw, rectangle, rounded corners=.3, thick, fill=black!2, drop shadow={shadow xshift=.5ex,shadow yshift=-.5ex}, font=\sf, minimum height=5ex,align=center},
  hwregblock/.style={hwblock, fill=blue!10},
  every circuit symbol/.style={hwblock},
  filteradd/.style={hwblock, circle, minimum height=1ex},
  filtermult/.style={hwblock, minimum height=7ex, regular polygon, regular polygon sides=3, shape border rotate=180, inner sep= .2ex},
  hwbus/.style={very thick,>={Stealth[length=6pt]}},
  hwbuslarge/.style={line width=2pt,>={Stealth[length=7pt]}},
  hwbuswhitebackground/.style={line width=4pt,color=white},
  hwwire/.style={thin, >={Stealth[length=4pt]}  },
  hwword/.style={draw, rectangle, minimum height=3ex},
  bitwidth/.style={font=\scriptsize,blue},
  abitwidth/.style={bitwidth,above},
  bbitwidth/.style={bitwidth,below},
  lbitwidth/.style={bitwidth,left},
  rbitwidth/.style={bitwidth,right},
  flopocoClass/.style={font=\small\tt,draw,align=center,rectangle,fill=blue!5,minimum height=3ex},
  flopocoInfo/.style={font=\small,align=center,rectangle, fill=yellow!50,fill opacity=0.5, text opacity=1,rounded corners=10}
}
\definecolor{normalbitcolor}{rgb}{0,0,0.5}
\definecolor{normalbitfillcolor}{rgb}{0.8,1,0.8}
\definecolor{signbitcolor}{rgb}{0.5,0,0}
\definecolor{signbitfillcolor}{rgb}{1,0.8,0.8}

\newboolean{showbitwidths} 
\setboolean{showbitwidths}{true}
\newcommand{\bitwidth}[3]{ 
  \ifthenelse{\boolean{showbitwidths}}{
    \draw[bitwidth,ultra thin] (#1) ++(0.5ex, 0.5ex)  -- ++(-1ex, -1ex) ++(0.5ex, 0.5ex)  node[#3bitwidth] {#2};
  }{}
} %

\newcommand{\drawbit}{rectangle +(-1,1)}   

\newcommand{\fixpointdot}[1] 
{
	\draw[black,fill] (0,#1-0.3) circle [radius=0.1];
}

\newcommand{\fixpointnumbernodot}[4] 
{ 
	\foreach \i in {#3,...,#2} {
		\draw (-\i,#4-0.3) \drawbit[normalbitcolor,fill=normalbitfillcolor];
	}
	\ifthenelse{#1=1}{
		\draw (-#2,#4-0.3) \drawbit[signbitcolor,fill=signbitfillcolor];
	} 
	{
		\draw (-#2,#4-0.3) \drawbit[normalbitcolor,fill=normalbitfillcolor];
	}
}

\newcommand{\fixpointbox}[4] 
{ 
   \draw[normalbitcolor,fill=normalbitfillcolor] (-#3,#4-0.3) rectangle +(#3-#2-1,1);
 }

\newcommand{\fixpointboxtrunc}[5] 
{ 
   \draw[normalbitcolor,fill=normalbitfillcolor] (-#5,#4-0.3) rectangle +(#5-#2-1,1);
   \draw[normalbitcolor,fill=normalbitfillcolor,opacity=0.5,fill opacity=0.5] (-#3,#4-0.3) rectangle +(#3-#5,1);
}
  \newcommand{\fixpointnumber}[4] 
{ 
	\fixpointnumbernodot{#1}{#2}{#3}{#4}
	\fixpointdot{#4};
}

\newcommand{\fixpointreal}[4] 
{ 
	\fixpointnumber{#1}{#2}{#3}{#4}
	\foreach \i in {0,...,9} 
	{ 
		\tikzmath{ \shade=\i*10;} 
		 \draw  (-#3+10-\i,#4-0.3)\drawbit[normalbitfillcolor!\shade,fill=normalbitfillcolor!\shade] ;
	}
		
}

\newcommand{\fixpointpositionsaxis}[4] 
{	
  }
  
\newcommand{\fixpointpositions}[4] 
{
  \fixpointpositionsaxis{#1}{#2}{#3}{#4}
	\ifthenelse{#1=1}{
	}{}
	\foreach \i in {#3,...,#2} {
		\draw[blue] (-\i-0.5,#4-1) node{\footnotesize \i};
	}
  \draw[blue] (-#3-0.5,#4-1) node[right=0.3]{\footnotesize $=\ell$};    
  \draw[blue] (-#2-0.75,#4-1) node[left]{\footnotesize $m=$};
  \draw[blue] (-#2-3, #4-1)  node[left] {\small bit exponent};
}

\newcommand{\integerbitpositions}[4] 
{
  \fixpointpositionsaxis{#1}{#2}{#3}{#4}
	\ifthenelse{#1=1}{
		\draw (-#2,#4) +(-0.5,0.2) node{\footnotesize $s$};
	}{}
	\foreach \i in {#3,...,#2} {
		\draw[blue] (-\i-0.5,#4-1) node{\footnotesize \i};
	}
  \draw[blue] (-#2-3, #4-1)  node[left] {\small bit exponent};
}

  \newcommand{\fixpointweights}[4] 
{	
	\ifthenelse{#1=0}{
		\draw (-#2-0.5,#4+1.5) node {\small $2^{#2}$};
	}{ 
    \draw (-#2-1, #4+1.5) node (w) {\small $-2^{#2}$~} ;
	}
\draw (-#3-0.5,#4+1.5) node {\small $2^{#3}$}; 
  \draw (-#2-3, #4+1.5)  node[left] {\small bit weight};

}

\newcommand{\fixpointallweights}[4] 
{
  \pgfmathtruncatemacro{\lsb}{#3} 
  \pgfmathtruncatemacro{\msb}{#2-#1} 
	\ifthenelse{#1=0}{
		\draw (-#2-0.5,#4+1.5) node {\small $2^{#2}$};
	}{ 
    \draw (-#2-1, #4+1.5) node (w) {\small $-2^{#2}$~} ;
	}
  \foreach \x in {\lsb,...,\msb} {
    \draw (-\x-0.5,#4+1.5) node {\small $2^{\x}$};
    }
  \draw (-#2-3, #4+1.5)  node[left] {\small bit weight};
}

\newcommand{\fixpointweightsml}[4] 
{	
	\ifthenelse{#1=0}{
		\draw (-#2-0.5,#4+1.5) node {\small $2^{m}$};
	}{ 
    \draw (-#2-1, #4+1.5) node (w) {\small $-2^{m}$~} ;
		\draw (-#2+1, #4+1.5) node (w) {\small $~~2^{m-1}$};
	}
	\draw (-0.5, #4+1.5) node {\small $2^0$};
  \ifthenelse{#3=0}{ 
    }{  \draw (-#3-0.5,#4+1.5) node {\small $2^{\ell}$}; }
  \draw (-#2-3, #4+1.5)  node[left] {\small bit weights};
}

\newcommand{\fpsignificand}[4] 
{
  \pgfmathsetmacro{\xinit}{#4} 
  \pgfmathsetmacro{\lsb}{#2} 
   \pgfmathtruncatemacro{\msb}{#2+#1} 
   \draw[normalbitcolor,fill=normalbitfillcolor] (-\lsb,#3-0.3) rectangle +(-#1-1,1);
	 \foreach \i [remember=\x as \xlast (initially \xinit)] in {\lsb,...,\msb} {
     \pgfmathtruncatemacro{\t}{\xlast / 2}
     \pgfmathtruncatemacro{\currentbit}{\xlast - 2*\t}
     \pgfmathsetmacro{\x}{\t}
     \draw (-\i-0.5,#3+0.2) node{\footnotesize \currentbit};
   }
}
\newcommand{\fpaccbits}[4] 
{
  \pgfmathsetmacro{\xinit}{#4} 
  \pgfmathsetmacro{\lsb}{#2} 
   \pgfmathtruncatemacro{\msb}{#2+#1} 
   \draw[draw=none,normalbitcolor,fill=normalbitfillcolor] (-\lsb,#3-0.3) rectangle +(-#1-1,1);
	 \foreach \i [remember=\x as \xlast (initially \xinit)] in {\lsb,...,\msb} {
     \pgfmathtruncatemacro{\t}{\xlast / 2}
     \pgfmathtruncatemacro{\currentbit}{\xlast - 2*\t}
     \pgfmathsetmacro{\x}{\t}
     \draw (-\i-0.5,#3+0.2) node{\footnotesize \currentbit};
   }
}

\newcommand{\fixpointvalue}[4] 
{
  \pgfmathsetmacro{\xinit}{#4} 
  \pgfmathsetmacro{\lsb}{#2} 
   \pgfmathtruncatemacro{\msb}{#1} 
	 \foreach \i [remember=\x as \xlast (initially \xinit)] in {\lsb,...,\msb} {
     \pgfmathtruncatemacro{\t}{\xlast / 2}
     \pgfmathtruncatemacro{\currentbit}{\xlast - 2*\t}
     \pgfmathsetmacro{\x}{\t}
     \draw (-\i-0.5,#3+0.2) node{\footnotesize \bf \currentbit};
   }
}

\newcommand{\fixpointrandomreal}[3] 
{	
	\fixpointdot{#3}
	\foreach \i in {#2,...,#1} {
		\ifthenelse{\i=#1}		{
			\draw (-\i-0.5,#3+0.2) node{\tt \footnotesize 1};
		}		{
			\draw (-\i-0.5,#3+0.2) node{\tt \footnotesize \pgfmathparse{isodd(random(2))}\pgfmathresult};
		}
	}
	
}

\newcommand{\FloatPlotZero} 
{
	\draw[red] (0,.15) -- ++(0,-.4) node[below] {\textcolor{red}{0}};	
}

\newcommand{\floatformatnotext}[4]{ 
    \pgfmathsetmacro{\posy}{-3.5*#1} 
    \pgfmathsetmacro{\mywE}{2*#2}  
    \pgfmathsetmacro{\mywF}{2*#3} 
    \foreach \x in {-#2,...,#3} {
      \draw[black!30](2*\x,\posy) -- ++(0,2);
    }
    \draw[fill=magenta, fill opacity=0.3](0,\posy) rectangle ++(-\mywE,2);
    \draw[fill=signbitfillcolor, fill opacity=0.3](-\mywE,\posy) rectangle ++(-2,2);
    \draw[fill=normalbitfillcolor, fill opacity=0.3](0,\posy) rectangle ++(\mywF,2);
    \draw (-\mywE,\posy) ++ (-3,1) node[left]{\small \bf #4};
  }
\newcommand{\floatformat}[4]{ 
    \pgfmathsetmacro{\posy}{-3.5*#1} 
    \pgfmathsetmacro{\mywE}{2*#2}  
    \pgfmathsetmacro{\mywF}{2*#3} 
    \foreach \x in {-#2,...,#3} {
      \draw[black!30](2*\x,\posy) -- ++(0,2);
    }
    \draw[fill=magenta, fill opacity=0.3](0,\posy) rectangle ++(-\mywE,2) node[opacity=1,midway]{\small $E$};
    \draw[fill=signbitfillcolor, fill opacity=0.3](-\mywE,\posy) rectangle ++(-2,2) node[opacity=1,midway]{\small $s$};;
    \draw[fill=normalbitfillcolor, fill opacity=0.3](0,\posy) rectangle ++(\mywF,2) node[opacity=1,midway]{\small $F$};;
    \draw (-\mywE,\posy) ++ (-3,1) node[left]{\small \bf #4};
  }

\newcommand{\FloatPlotNormals}[4] 
{
  \pgfmathsetmacro{\wE}{#1} ; 
  \pgfmathsetmacro{\wF}{#2} ; 
  \pgfmathtruncatemacro{\minexp}{#3};  %
  \pgfmathtruncatemacro{\maxexp}{#4}; 
	\pgfmathsetmacro{\pwE}{exp(ln(2)*\wE)+.5}; 
	\pgfmathtruncatemacro{\pwF}{exp(ln(2)*\wF)+.5}; 

  \foreach \e in {\minexp,...,\maxexp} {
    \pgfmathtruncatemacro{\xscaled}{\pwE*(exp(ln(2)*\e))+0.5}; 
    \pgfmathsetmacro{\x}{\xscaled/\pwE}; 
    \draw[thin] (\x,-.15) -- ++(0, .5);
    \draw[thin] (-\x,-.15) -- ++(0, .5);
    
    \pgfmathtruncatemacro{\pwFm}{\pwF-1}; 
    \foreach \f in {1,...,\pwFm} {
    \pgfmathsetmacro{\xu}{\x+\f*\x/\pwF}; 
    \draw[thin] (\xu,-.15) -- ++(0, .3);
    \draw[thin] (-\xu,-.15) -- ++(0, .3);
    }
  }
}

\newcommand{\FloatPlotIEEENormals}[2] 
{
  \pgfmathsetmacro{\wE}{#1} ; 
  \pgfmathsetmacro{\wF}{#2} ; 
  \pgfmathtruncatemacro{\pwE}{exp(ln(2)*\wE)+.5}; 
  \pgfmathtruncatemacro{\minexp}{2-\pwE/2};  %
  \pgfmathtruncatemacro{\maxexp}{\pwE/2-1};
  \FloatPlotNormals{\wE}{\wF}{\minexp}{\maxexp}
}

\newcommand{\FloatPlotIEEESubnormals}[2] 
{
  \pgfmathsetmacro{\wE}{#1} ; 
  \pgfmathsetmacro{\wF}{#2} ; 
  \pgfmathtruncatemacro{\pwE}{exp(ln(2)*\wE)+.5}; 
  \pgfmathtruncatemacro{\minexp}{2-\pwE/2};  %
  \pgfmathtruncatemacro{\pwFm}{\pwF-1}; 
  \pgfmathtruncatemacro{\xscaled}{2*\pwE*(exp(ln(2)*\minexp))+0.5}; 
  \pgfmathsetmacro{\x}{\xscaled/(2*\pwE)}; 

  \foreach \f in {0,...,\pwFm} {
    \pgfmathsetmacro{\xu}{\f*\x/\pwF}; 
    \draw[thick,red] (\xu,-.15) -- ++(0, .3);
    \draw[thick,red] (-\xu,-.15) -- ++(0, .3);
    }
}

\newcommand{\FloatPlotFlopoco}[2] 
{
  \pgfmathsetmacro{\wE}{#1} ; 
  \pgfmathsetmacro{\wF}{#2} ; 
  \pgfmathtruncatemacro{\pwE}{exp(ln(2)*\wE)+.5}; 
  \pgfmathtruncatemacro{\pwF}{exp(ln(2)*\wF)+.5}; 
  \pgfmathtruncatemacro{\minexp}{-\pwE/2+1};  %
  \pgfmathtruncatemacro{\maxexp}{\pwE/2};
  \FloatPlotNormals{\wE}{\wF}{\minexp}{\maxexp}
  \draw[red,thin] (0,-.3) -- (0, .4);
}

\usepackage{xcolor}

\colorlet{greenbg}{blue!10!green!30!white}
\colorlet{greentext}{blue!10!green!50!black}
\colorlet{yellowbg}{yellow!20!white}
\colorlet{yellowbg2}{yellow!50!white}
\colorlet{yellowtext}{yellow!10!black}
\colorlet{redbg}{red!20!white}
\colorlet{redtext}{red!30!black}

\usepackage[
  colorinlistoftodos, 
  textwidth=\marginparwidth, 
  textsize=scriptsize, 
  ]{todonotes}

\graphicspath{{fig/}}

\newcommand{\RR}{\mathbb{R}}
\newcommand{\ZZ}{\mathbb{Z}}

\newcommand{\abs}[1]{\left\lvert #1 \right\rvert}

\newcommand{\floor}[1]{\left\lfloor #1 \right\rfloor}
\newcommand{\ceil}[1]{\left\lceil #1 \right\rceil}

\newcommand{\Sum}{\sum\limits}
\newcommand{\Min}{\min\limits}

\makeatletter
\newcommand{\ostar}{\mathbin{\mathpalette\make@circled\star}}
\newcommand{\make@circled}[2]{%
  \ooalign{$\m@th#1\smallbigcirc{#1}$\cr\hidewidth$\m@th#1#2$\hidewidth\cr}%
}
\newcommand{\smallbigcirc}[1]{%
  \vcenter{\hbox{\scalebox{0.77778}{$\m@th#1\bigcirc$}}}%
}
\makeatother

\newcommand{\xlo}{\underline{x}}
\newcommand{\xhi}{\overline{x}}
\newcommand{\ylo}{\underline{y}}
\newcommand{\yhi}{\overline{y}}
\newcommand{\zlo}{\underline{z}}
\newcommand{\zhi}{\overline{z}}

\newcommand{\wE}{w_E}
\newcommand{\wF}{w_F}

\newtheorem{defi}{Definition}

\newtheorem{prop}{Property}

\newcommand{\F}{\textsc{Faust} }

\title{Towards fixed-point formats determination\\ for \F programs}

\author{Agathe Herrou\text{*}\\ herrou@grame.fr\thanks{\text{*} Grame-CNCM, INSA Lyon, Inria, CITI, UR3720, 69621 Villeurbanne, France} 
  \and Florent de Dinechin\textdagger \\ florent.de-dinechin@insa-lyon.fr\thanks{\textdagger INSA Lyon, Inria, CITI, UR3720, 69621 Villeurbanne, France}
  \and Stéphane Letz\text{*} \\ letz@grame.fr
  \and Yann Orlarey\textdaggerdbl  \\ yann.orlarey@inria.fr\thanks{\textdaggerdbl Inria, INSA Lyon, CITI, UR3720, 69621 Villeurbanne, France}
  \and Anastasia Volkova\textdaggerdbl \\ anastasia.volkova@inria.fr
}

\begin{document}
\maketitle
\begin{abstract}
  Modern programmable digital signal processing relies on floating-point numbers for their ease of use.
  Fixed-point number formats have the potential to save resources and improve execution time,
  but realising this potential burdens the programmer with the need to define each format, at every step of the computation.
  This article reviews existing methods to automatically determine fixed-point formats, then describes and evaluates the prototype implementation of automatic fixed-point format determination in the \F compiler.
\end{abstract}


\section{Introduction}
This article discusses the use of fixed-point numbers to replace floating-point numbers in audio programs, particularly in the context of the \F compiler.

\F is a functional programming language for sound synthesis and digital signal processing.
The \F compiler~\cite{faustcompiler} generates low-level target code
(by default C++, but over the years it has come to support several dozen targets)
from a high-level specification written in the \F language.

A recent development in the \F ecosystem is the Syfala~\cite{syfala} toolchain which brings a new target: FPGA-based boards (Field Programmable Gate Arrays).
An FPGA is an integrated circuit designed to emulate arbitrary digital circuits.
It offers an array of configurable boolean gates which can be linked by a  configurable network of wires.
FPGAs are thus user-programmable, but where the processor programming model is a sequence of instructions, the programming model of FPGAs is the digital circuit: a graph of gates and registers at the bit level.
Running audio programs on FPGAs enables very low latency, down to single-sample buffers.
However, programming FPGAs involves the arduous task of designing the circuit, and the objective of Syfala is to compile \F programs directly into FPGA circuits.

In this alternative programming model, it is possible to reconsider the way real numbers are represented and transformed.
Audio programs meant to be run on general-purpose processors typically use floating-point numbers,
an approximation of real numbers which is very versatile and thus convenient to programmers.
Another way of representing real numbers in a computer is to use fixed-point formats, where numbers are represented as integers with a constant scale.
Fixed-point operations reduce to integer operations with little or no overhead.
Table~\ref{tab:hardware_fl_fp} shows that operations on integers are significantly less expensive than operations on floating-point numbers on FPGAs.
Furthermore, the inputs and outputs of an audio pipeline (entering the analog-to-digital and leaving the digital-to-analog converters) are not floating-point numbers, but fixed-point numbers.

\begin{table}
\begin{center}
\begin{tabular}{|r|r|r|}
\hline
  Operation             & area          & delay                         \\
  \hline
  \hline
  Float32 $+$           & 313 LUT       & 11.4 ns                       \\
  Int32 $+$             & 32 LUT        & 1.8 ns                        \\
  Int24 $+$             & 24 LUT        & 1.7 ns                        \\
  \hline
  Float32 $\times$      & 2 DSP, 66 LUT & 6.8 ns                        \\
  Int32 $\times$        & 4 DSP, 47 LUT & 5.8 ns                        \\
  Int24 $\times$        & 2 DSP, 0 LUT  & 4.5 ns                        \\
  \hline
\end{tabular}
\end{center}
\caption{Silicon area use (expressed in number of Look-Up Tables and Digital Signal Processing blocks) and input-to-output delay comparison on AMD Zynq FPGAs (data obtained with FloPoCo~\cite{flopoco} operators).}
\label{tab:hardware_fl_fp}
\end{table}

Most processors offer hardware support for a handful of standard formats: 8, 16, 32 and 64-bit integers, and in floating-point the 32-bit \texttt{float} and the 64-bit or \texttt{double} formats.
On FPGAs, formats can  be adjusted more finely.
However this is a difficult and error-prone task.
The present work aims at giving the \F compiler automatic capabilities for inferring fixed-point formats in a program in a way that preserves audio quality.

This article mostly targets fixed-point for FPGAs, but the tools developed here are also useful in providing a better understanding of the numerical behaviour of \F programs, irrespective of the target.
We will lay out the basics of numerical formats in Section~\label{sec:formats},
describe the principles of fixed-point format determination in Section~\label{sec:determination},
explain how these principles have been implemented into the Faust compiler in Section~\label{sec:inference},
and finally present the results of these algorithms on two real-life Faust programs in Section~\label{sec:results}.

\section{Basics of numerical formats}\label{sec:formats}
Mainstream number formats in computers have in common that they are encoded in binary in a finite, fixed number of bits (usually 8, 16 or 32 bits for audio applications).

\subsection{Binary versus decimal}
Binary is not fundamentally different from decimal, but some numbers finitely representable in decimal have an infinite representation in binary.
For instance, $0.1$ (one tenth) is written in binary $0.0001100110011001100... = 0.000(1100)^\infty$, therefore the computer will not store exactly a number input in decimal as 0.1.
Conversely, numbers with an infinite decimal representation, such as $1/3$ or $\pi$, also have an infinite binary representation.



\subsection{Floating-point numbers}

The most frequently used representation of real numbers are floating-point numbers.
The main advantage of this representation is a very large range of representable values, covering astronomical as well as infinitesimal values. Because of that, it is a versatile  ``one-size-fits-all'' solution,
hence the representation of choice for programs involving reals.

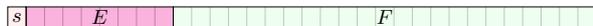
\begin{figure}[h]
  \centerline{
    \scalebox{0.85}{
     \begin{tikzpicture}[x=.9ex,y=1ex];
       \floatformat{0}{8}{23}{}
     \end{tikzpicture}
     }
  }
\caption{Representation of single-precision floating-point numbers: a sign bit, 8 exponent bits, 23 fraction bits.}
\label{fig:fl_rep}
\end{figure}

Technically, computer floating point is the binary version of the decimal so-called scientific notation (which represents the speed of light as $c\approx 3.00 \times 10^8$ and the mass of the proton as $m\approx 1.67 \times 10^{-27}$ kg).
So-called normal floating-point numbers are of the form $(-1)^s\times 1.F\times 2^E$ where $s$ is a sign bit, and both  the exponent $E$ and the fraction $F$ are encoded in binary on a fixed number of bits (Figure~\ref{fig:fl_rep}).
Two values of the exponent bits are reserved to encode special values: infinities, subnormals and zero -- for details see \cite{Goldberg91} or \cite{MullerEtAl2018:HandBook}.
The distribution of floating-point values is illustrated (in a toy 6-bits example) in Figure~\ref{fig:FPvsFixP}.




A floating-point format is defined by the sizes (in bits) of the fraction and exponent fields (Figure~\ref{fig:fl_rep}).
The IEEE-754~\cite{ieee754} standard defines these parameters for standard formats
including the FP32 and FP64 formats available in C/C++ under the names \texttt{float} and \texttt{double}.

The IEEE-754 format also specifies the behaviour of standard operations.
This will be described in Section~\ref{sec:format:operations}.
Implementing these operations in hardware is costly, both in terms of circuit size and of execution time.
On general-purpose hardware, the versatility of the format makes up for the hardware cost.
When compiling a \F program into a circuit, a simpler fixed-point format may be more efficient.

\subsection{Fixed-point numbers}\label{sec:fx}



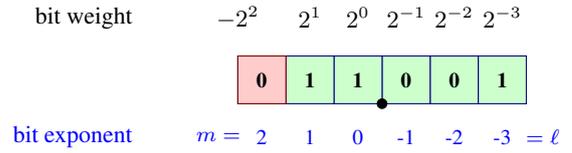
\begin{figure}
\centerline{
  \begin{tikzpicture}[x=4ex, y=4ex]
      \fixpointnumber{1}{2}{-3}{0}
      \fixpointpositions{1}{2}{-3}{0}
      \fixpointallweights{1}{2}{-3}{0}
      \fixpointvalue{2}{-3}{0}{25}
    \end{tikzpicture}
  }
    \caption{$\pi$ rounded to the toy fixed-point example of Figure \ref{fig:FPvsFixP}: $\pi\approx 3.125=2^1+2^0+2^{-3}$.}
    \label{fig:fx_rep-toy}
\end{figure}

A fixed-point number is a vector of bits $(x_m, x_{m-1}, \ldots, x_\ell)$ which represents the signed rational value
\begin{equation}\label{eq:twos_complement}
  X = -2^mx_m + \Sum_{i=\ell}^{m-1}  2^ix_i 
\end{equation}
where $m$ and $\ell$ are two integers that define the exponents of the most significant bit (MSB) and least significant bit (LSB) respectively.
An example of a toy 6-bit fixed-point format with $m=2$ and $\ell=-3$ is shown in Figure~\ref{fig:fx_rep-toy}.
As this figure shows, bits with a negative exponent are to the right of the fractional point.
Another interpretation is that Equation~\eqref{eq:twos_complement} describes a $w$-bit binary integer (with $w=m-\ell+1$) scaled by $2^\ell$. 

In this work, all fixed-point numbers are signed thanks to two's complement representation: the leftmost bit has a negative weight $2^m$, and is called the sign bit.
Two's complement may be less intuitive than a sign-magnitude representation (as in floating point), but it is the format of choice in binary arithmetics because it is very hardware-friendly \cite{asa}.

\begin{figure*}[t]
	\centering
  \begin{tikzpicture}[x=6ex, y=5ex]
    \pgfmathsetmacro{\yfix}{2} ; 
    \pgfmathsetmacro{\xcut}{5.2} ; 

    \begin{scope}[x=1.5ex, y=1.5ex,xshift=-40ex,yshift=13ex] 
      \fixpointnumber{1}{2}{-3}{0}
    \end{scope}
    \foreach \x in {-32,...,31} {
      \draw[black](0.125*\x,\yfix-.15) -- ++(0,.3);
    }

    \draw[red] (3.1415926,-.4)node[below]{$\pi$} -- ++(0,3) ;

    \draw (-\xcut-2,\yfix) -- (\xcut+4,\yfix) node[below] {$\mathbb{R}$};
    \draw(3.875, \yfix+.7) node[] {$\frac{31}{8}$};
    \draw(-4, \yfix+.7) node[] {$-\frac{32}{8}$};

    \begin{scope}[x=.75ex, y=.75ex,xshift=-39ex,yshift=-3ex] 
      \floatformatnotext{0}{3}{2}{}
    \end{scope}
    \draw[] (-\xcut,0) -- (\xcut,0);
    \draw[dotted](-\xcut-1,0) -- (\xcut+1,0);
    \draw (\xcut+1.2,-.15) -- ++(0,.3);
    \draw(\xcut+1.2, .7) node[] {12};
    \draw (\xcut+3.2,-.15) -- ++(0,.3);
    \draw(\xcut+3.2, .7) node[] {14};
    \draw(\xcut+4.2, .7) node[] {$+\infty$};
    \draw (\xcut+1,0) -- (\xcut+4,0) node[below] {$\mathbb{R}$};
    \draw (-\xcut-1,0) -- ++(-1.5,0);
    \draw (-\xcut-1.5,-.15) -- ++(0,.3);
    \draw(-\xcut-1.5, .7) node[] {-14};
    \draw(-\xcut-2.5, .7) node[] {$-\infty$};

    \draw[thin] (0, -.3)node[below]{$0$} --(0,\yfix) --+(0,.5);

    \draw(.5, .7) node[] {$\frac{1}{2}$};
    \draw(1, .7) node[] {1};
		\draw(2, .7) node[] {2};
		\draw(4, .7) node[] {4};

    \clip[] (-\xcut,-1) rectangle (\xcut,1); 
    \FloatPlotIEEENormals{3}{2}
    \FloatPlotIEEESubnormals{3}{2}

  \end{tikzpicture}
  \caption{Representation on the real axis of the numbers representable in two toy 6-bit formats: above, two's complement fixed-point with 3 integral bits and 3 fraction bits. Below, floating-point with 3 exponent bits and 2 fraction bits (longer bars are powers of two, subnormals are in red, maximum finite representable value is $1.11_2\cdot 2^{3}=14$, two infinities). }
	\label{fig:FPvsFixP}
\end{figure*}
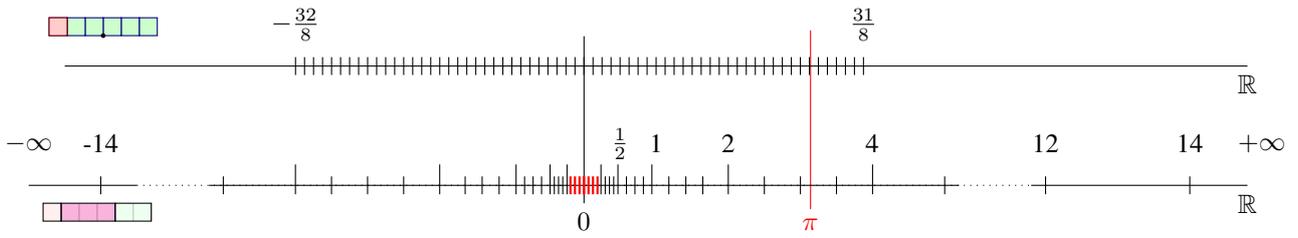

The range of values representable on a fixed-point number with parameters $(m, \ell)$ is $[-2^{m-1}; 2^{m-1}- 2^\ell]$, with a constant step of $2^\ell$, as shown in Figure~\ref{fig:FPvsFixP}.
The parameter $m$ thus defines the range of the format, and the parameter $\ell$ defines its resolution or precision.

Such limitations of the formats illustrate the importance of tailoring the format of a variable to the values it will hold.
This is the reason why it is more difficult to design programs using fixed-point numbers:
a good format has to be inferred for every variable present in the program,
unlike with floating-points where it is generally safe to use generic formats throughout the program.

This being said, fixed-point also has the potential of providing more accuracy than floating-point for a given number of bits.
This is in particular true for constants (whose point, by definition, does not float), as illustrated by Figure~\ref{fig:FPvsFixP} for the constant $\pi$: with the toy formats used here it is obvious that the fixed-point value 3.125 is closer to $\pi$ than the closest floating-point value, here $1.5\cdot 2^1$.

\subsection{Operations}
\label{sec:format:operations}

In the general case, the exact result of an arithmetic operation or function (such as $+$, $\times$, $/$, $\sqrt{~}$ or cosine) on machine numbers inputs is not a machine number.
It may not fit in the representable range, which is called an overflow situation.
Alternatively, it may be in range but not be one of the discrete representable values, in which case it must be rounded.

\subsubsection{Managing overflow}\label{sec:overflow}
Floating-point formats will rarely overflow on practical audio applications (even single precision can represent numbers up to $10^{38}$).
If it were to happen nonetheless, these formats offer two infinities ($+\infty$ and $-\infty$) that capture overflow.
These are outside of the scope of this article.

In fixed point, there are several ways to handle overflow situations.
They can simply be ignored: the cheapest binary operators for addition and multiplication implement arithmetic modulo $2^m$, so for instance a value larger than $2^{m-1}$ will be wrapped to a negative value.
In the case of an audio signal, this will cause undesired discontinuities that ultimately result in audible distortion.
Another option is saturated arithmetic: in case of overflow (positive or negative), the largest (or largest negative) representable value is returned.
Saturated fixed-point arithmetic involves some hardware and latency overhead.

The best option, and the one we aim to develop, is to make sure that overflows will not happen, by using, for each variable, a large enough value of $m$.
For a constant $C$, this is easy: $m=\lfloor\log_2 (|C|)\rfloor+1$ will do (the $+1$ is for the sign bit).
For instance, $\log_2 (\pi) \approx 1.65$, gives $m=2$ (see Figure~\ref{fig:fx_rep-toy}).

For variables, $m$ will similarly be the $\log_2$ of the largest possible value the variable can take, which may be evaluated by \emph{interval range analysis}.
This will be the subject of section~\ref{sec:range}.

\subsubsection{Managing rounding}
\label{sec:format:rounding}

The exact result of an operation is not necessarily a representable number, in which case it must be rounded.
We will denote by $\circ(x)$ the rounding of number $x$.
Rounding the exact result to the nearest machine number is called correct rounding.
It is easy to compute for the four arithmetic operations ($+$, $-$, $\times$, $/$), and the IEEE-754 standard for floating point arithmetic mandates correct rounding.
For other operations or functions, correct rounding may be quite expensive to guarantee, but a slightly degraded approach is to return one of the two numbers surrounding the exact result.
This is called faithful rounding, and for reasons beyond the scope of this article, it is often much cheaper than correct rounding~\cite{asa}.
For instance $\pi$, on the 6-bit fixed-point format of Figure~\ref{fig:FPvsFixP}, $\pi$ could be rounded faithfully to 3.125 (the nearest result) or 3.250 (the value immediately above).
This example illustrates that rounding involves an error (none of these values is $\pi$) but also that the error of faithful rounding  is bounded by $2^\ell$, the resolution of the format (here $1/2^3=0.125$).
Here the error of rounding $\pi$ to 3.250 is $\delta=3.250-\pi = 0.1084...<0.125$.
Similarly, correct rounding entails an error bounded by $\frac{1}{2}2^\ell$.

The error can be made smaller by using a smaller value of $\ell$, which will entail a finer resolution $2^\ell$.
Determining a small enough value of $\ell$ will be the subject of  Section~\ref{sec:accuracy}.

In floating-point, the difference (visible in Figure~\ref{fig:FPvsFixP}) is that the resolution changes with the magnitude of the number.
The absolute rounding error $\circ(x)-x$ no longer has a constant bound for the format.
However, the relative rounding error $\frac{\circ(x)-x}{x}$ has a constant bound related to $2^p$, where $p$ is the number of bits of the significand (or mantissa).

\subsubsection{Operator implementation}

Given an arithmetic operator or real function, and input and output number formats,
the problem of designing an implementation that respect certain properties
({\em e.g.} faithfully rounded) comes up.
This problem is adressed by software such as FloPoCo~\cite{flopoco} or VFloat~\cite{vfloat},
which generate operators implementation from a specification,
or \lstinline{ap_fixed}~\cite{apfixed}, a C++ library of fixed-point formats for High-Level Synthesis of programs for FPGAs.

\subsection{Machine numbers and audio by example}\label{sec:audioexample}

To illustrate how the behaviour of a program using floating-point or fixed-point diverges from the expected behaviour with real numbers, let us take the example of a very simple program: a ramp.
A ramp can be generated with the following \F program:
\mbox{\lstinline{process = 0.5 : + ~ _;}}.\footnote{This code can be executed and tinkered with in the Faust online IDE: \href{https://faustide.grame.fr/}{\url{https://faustide.grame.fr/}}}
At each iteration, this program outputs a certain value, and feeds it back into a recursive loop where it is incremented by $0.5$ (a value exactly representable even in our toy 6-bit formats).
Eventually, the limits of the format used to represent the output will be attained, resulting in an overflow.

In the case of floating-point, the value saturates when the amplitude of the value gets too large relative to the increment.

For example, if the format only allows for two fraction bits like in Figure~\ref{fig:FPvsFixP}, 
as soon as the output value reaches~$4$, adding $0.5$ will not have any further effect as the exact sum $4.5$ is not representable (Figure~\ref{fig:FPvsFixP}) and will be rounded back to $4$ (it is exactly between $4$ and $5$, so we apply here the \emph{tie to even} rule of IEEE754~\cite{Goldberg91}).
What should be a ramp becomes a plateau.

If we replace the increment $0.5$ with $0.375$ (still representable in binary as $0.011_2$), the values computed by this program are $0$, $0.375$ (representable), $0.75$ (representable), $1$ (as $1.125$, not representable, must be rounded), $1.5$ ($1.375$ is not representable, and rounded up due to the tie to even rule), $2$, $2.5$, $3$, $3.5$, $4$, $4$, $4$\ldots What we observe is that due to rounding, the actual increment changes (from $0.375$ to $0.25$ then $0.5$) before the plateau is reached.

In the case of fixed-point, there will be no such irregular rounding effect.
There will be a plateau if the addition is saturated.
However, if the cheapest adder is used, it implements addition modulo $8$: the value $4$ (not representable) is transparently wrapped to $-4$, and the ramp proceeds from there.
What should be a ramp becomes a sawtooth signal (whose period is determined by the format).

Of course, it would be a very bad idea to feed such a ramp to a sine function in the hope to obtain a sine wave, although this makes perfect sense from a mathematical point of view.

\section{Fixed-point Format determination}\label{sec:determination}

In order to determine the fixed-point format most adapted to a given variable,
one has to determine both $m$ and $\ell$.

The value of $m$ is deduced from the amplitude of values attained by the variable.
The value of $\ell$ can either be deduced from a fixed bitwidth $w$ with the formula $w = m - \ell + 1$,
or has to be chosen so that the rounding errors and their propagation through the program remain under a certain threshold.

\subsection{Range of a value}\label{sec:range}

The exponent $m$ of the most significant bit has to be chosen so that the value stored in the variable does not overflow.
This can be done by performing range analysis on a program.

The range of the inputs to a program can generally be inferred immediately.
For example, an input audio signal will typically have its value constrained to $[-1;1]$; 
a value that is produced by a slider will have its minimum and maximum values specified as parameters of the slider;
constants have already been addressed in Section~\ref{sec:overflow}.

Range analysis consists in recursively inferring the ranges of values taken by the other variables.
Knowing the range of the input to an operation,
the output range can be straightforwardly deduced by computing the image of the input range through the operation.
Traditional interval analysis~\cite{Moore66} slightly enlarges this image to cater to any case of rounding.

For example, if the value input to an occurrence of function $\sin$
is in the range $[0; \pi]$,
then the output will be constrained to $\sin([0; \pi]) = [0;1]$.

Program loops are more tricky to address.
The ideal output of the previous ramp has the range $[0, +\infty)$.
Techniques such as abstract interpretation~\cite{fluctuat} have been designed to detect this in finite time, they are out of scope of this article.

More interesting is the case of infinite (or arbitrary large) loops where the variable have a finite range.
For instance, in the following simple pseudo-code program
\begin{lstlisting}[language=C]
x = 1;
while (true)
  x = x/2;
\end{lstlisting}
starting with the interval $[1,1]$,  applying the interval propagation rule, then taking the convex hull of the union of the previous and new range of the variable  \lstinline{x} converges to the range interval  $[0;1]$.

Unfortunately this kind of procedure cannot always converge to tight intervals.
Sometimes a human may prove the range of a variable only by using information that is inaccessible to a compiler, for instance a known correlation between inputs.
In such cases, the compiler must fall back to floating-point, or to fixed-point with some arbitrary value of $m$.

\subsection{Accuracy}\label{sec:accuracy}

The other element to infer is the position $\ell$ that should be given to the least significant bit.
If the width $w$ of the fixed-point number is fixed (as would be the case in processor without floating-point such as early DSP processors),
the computation of $\ell$ is immediate from the relation $w = m - \ell + 1$.
If there is no such constraint,
one has to determine a value of $\ell$ that matches the desired precision over the signal.

Every value present in a numerical program is bound to carry two types of errors.
Extrinsic errors are due to external factors such as measurement, and discretisation.
We will not deal with this kind of value in this work.
Intrinsic errors come from the finite-precision representation of the provided real value.
Each operation contributes a rounding error smaller than the weight of its output least significant bit (see Section~\ref{sec:format:rounding}) but these elementary rounding errors accumulate through the computation.

To capture this, the error is modelled by an interval $E$ enclosing the difference between the real value of the variable and its implemented value.
If the error interval $E$ has amplitude $\epsilon$,
then the bits located after position $\log_2(\epsilon)$ are not guaranteed to contain accurate information.

When a variable is input to a function or an operator,
its error will affect the output as well.
Let us consider a finite-precision value $\hat{x}$, approximating a real value $x$,
with $\epsilon$ the corresponding error defined by: $\hat{x} = x + \epsilon$.
Assuming that $\epsilon$ is several orders of magnitude smaller than $x$,
the value $f(\hat{x})$ can be approximated by $f(x) + \epsilon f'(x)$
through a first order Taylor expansion.
The term $\epsilon f'(x)$ represents the error committed over the output value
that is purely due to the input error:
the input error is magnified by the amplitude of $f'(x)$.
In order to balance this contribution to the error with the rounding error due to $f$,
the output precision should be chosen so that the weight of the least significant bit
is the same order of magnitude as $\epsilon f'(x)$.

From error analysis, constraints relating the input and output precisions of a function or operator can be deduced.
Supplementary constraints can be added, such as information on input/output precisions,
or precisions at other key points of the program.
For instance, if an expression uses an approximation, it makes little sense to compute much more accurately than this approximation. 
The set of all constraints present in a program are arranged over the signal graph of the program
and needs to be solved in order to determine formats to use in the program.

If the signal graph is acyclical, then a simple transversal is enough to propagate the constraints and solve them.
If the graph has cycles (that is if the program, like the one presented in Section~\ref{sec:audioexample}, presents a feedback loop),
then there is a risk that the interval evaluation of accumulated errors diverges.
However, there are large classes of problems, for instance Linear Time-Invariant filters, where ad-hoc techniques can be used to tightly capture the propagation of ranges and errors \cite{VIDH2019}.
In the general case, there will always remain classes of loops for which no satisfying analytical solution can be found.
For these loops, a fallback solution has to be defined.
Our solution of choice is to set a default precision for signals in the feedback of loops,
which we arbitrarily chose to be $-24$.
We currently deal with all loops this way,
but we plan to implement techniques adapted to LTI filters in future versions of this work.

\subsection{Tools for code analysis and synthesis}

A number of tools have been proposed to assist in the design of fixed-point programs.
Some are fully automatic but restricted in scope \cite{minibit}, some are more widely applicable but require user interaction \cite{fridge,gappa,rosa,daisy}.
Some rely on profiling~\cite{kim1998}, but most rely on worst-case analysis.
Among the latter, most use interval arithmetic, sometimes with extensions such as affine arithmetic~\cite{minibit,fluctuat}, Taylor models~\cite{daisy} and SMT solving~\cite{kinsman2010,rosa,daisy} to compute tight intervals for composite expressions.
Some start with a C/C++ program~\cite{fridge,fluctuat}, some start with a higher-level description of the problem, e.g. using real numbers and context constraints~\cite{rosa,daisy,gappa}.
Some focus on the analysis of programs~\cite{kim1998,fluctuat,gappa}, some generate fixed-point implementations \cite{minibit,fridge}.

A good recent survey is part II of~\cite{approxcomputing}, and the Daisy project~\cite{daisy} aims at gathering state-of-the-art techniques in one open-source framework.

\section{Fixed-point format inference in \F}\label{sec:inference}

\F is a domain-specific language for audio programming,
but can nonetheless express a class of program that goes beyond LTI filters.
Thus, an automatic fixed-point format inference system for \F
can not fully rely on assumptions on the nature of programs.
In this section, we lay out our precision inference method and its integration in the \F compiler.

\subsection{The \F Compiler}

\begin{figure*}
  \begin{subfigure}[b]{\textwidth}
    \centerline{
        \scalebox{0.6}{
  \begin{tikzpicture}[x=\tkzscalex,y=\tkzscaley]
  
  \node[draw=greentext, text = greentext, circle, fill=greenbg, minimum height = 2*\tkzscaley,text width=2*\tkzscalex,align=center,] (FAUST) at (0,0) {Faust Code};
  
  \node[draw=yellowtext, text=yellowtext, rounded corners=2pt, fill=yellowbg2, text width=2*\tkzscalex, align=center, minimum height = 2*\tkzscaley, right = 0.5*\tkzscalex of FAUST] (SEMANTIC)  {Semantic Phase};

  \node[text=greentext, draw= greentext, fill=greenbg, ellipse, text width=2.5*\tkzscalex, align=center, minimum height = 2*\tkzscaley, right = 0.5*\tkzscalex of SEMANTIC, inner sep = 2pt] (SIGNAL)  {List of signals in Normal Form};
  
  \node[draw=yellowtext, text=yellowtext, rounded corners=2pt, fill=yellowbg2, text width=2*\tkzscalex, align=center, minimum height = 2*\tkzscaley, right = 0.5*\tkzscalex of SIGNAL] (GENERATION) {Code Generation Phase};
  
  \node[text=redtext, draw= redtext, fill=redbg, ellipse, text width=2.5*\tkzscalex, align=center, minimum height = 2*\tkzscaley, right = 0.5*\tkzscalex of GENERATION, inner sep = 2pt] (IMPLEMENTATION)  {Implemen-tation: C++, js, LLVM\ldots};

  \draw[->] (FAUST) -- (SEMANTIC);
  \draw[->] (SEMANTIC) -- (SIGNAL);
  \draw[->] (SIGNAL) -- (GENERATION);
  \draw[->] (GENERATION) -- (IMPLEMENTATION);
  
  \scoped[on background layer]{\node [fit=(SEMANTIC)(SIGNAL)(GENERATION), rounded corners = 2pt, fill=yellowbg, draw=yellowtext, inner sep = 5pt] {};}
  
\end{tikzpicture}
}
      }
    \caption{The \F compilation chain.}
    \label{fig:faustchain}
  \end{subfigure}
  
  \begin{subfigure}[b]{\textwidth}
    
    \centerline{
        \scalebox{0.6}{
  \begin{tikzpicture}[x=\tkzscalex,y=\tkzscaley]
  
  \node[draw=greentext, text = greentext, circle, fill=greenbg, minimum height = 2*\tkzscaley,text width=2*\tkzscalex,align=center] (FAUST) at (0,0) {Faust Code};
  
  \node[draw=yellowtext, text=yellowtext, rounded corners=2pt, fill=yellowbg2, text width=2*\tkzscalex, align=center, minimum height = 2*\tkzscaley, right = 0.5*\tkzscalex of FAUST] (CALCULUS)  {Evaluation ($\lambda$-calculus)};

  \node[text=greentext, draw= greentext, fill=greenbg, ellipse, text width=2.5*\tkzscalex, align=center, minimum height = 2*\tkzscaley, right = 0.5*\tkzscalex of CALCULUS, inner sep = 2pt] (CIRCUIT)  {Circuit};
  
  \node[draw=yellowtext, text=yellowtext, rounded corners=2pt, fill=yellowbg2, text width=2*\tkzscalex, align=center, minimum height = 2*\tkzscaley, right = 0.5*\tkzscalex of CIRCUIT] (PROPAGATION) {Symbolic Propagation};
  
  \node[text=greentext, draw= greentext, fill=greenbg, ellipse, text width=2.5*\tkzscalex, align=center, minimum height = 2*\tkzscaley, right = 0.5*\tkzscalex of PROPAGATION, inner sep = 2pt] (SIGNAL)  {List of Signals};
  
  \node[draw=yellowtext, text=yellowtext, rounded corners=2pt, fill=yellowbg2, text width=2*\tkzscalex, align=center, minimum height = 2*\tkzscaley, right = 0.5*\tkzscalex of SIGNAL] (NORM) {Simplifica-tion and Normalisation};
  
  \node[text=greentext, draw= greentext, fill=greenbg, ellipse, text width=2.5*\tkzscalex, align=center, minimum height = 2*\tkzscaley, right = 0.5*\tkzscalex of NORM, inner sep = 2pt] (SIGNAL2)  {List of signals in Normal Form};

  \draw[->] (FAUST) -- (CALCULUS);
  \draw[->] (CALCULUS) -- (CIRCUIT);
  \draw[->] (CIRCUIT) -- (PROPAGATION);
  \draw[->] (PROPAGATION) -- (SIGNAL);
  \draw[->] (SIGNAL) -- (NORM);
  \draw[->] (NORM) -- (SIGNAL2);
  
  \scoped[on background layer]{\node [fit=(CALCULUS)(CIRCUIT)(PROPAGATION)(SIGNAL)(NORM), rounded corners = 2pt, fill=yellowbg, draw=yellowtext, inner sep = 5pt] {};}
    
\end{tikzpicture}
}
      }
    \caption{The semantic phase of the compilation.}
    \label{fig:semanticphase}
  \end{subfigure}
  \caption{Overview of the structure of the \F compiler.}
\label{fig:faustcompiler}
\end{figure*}
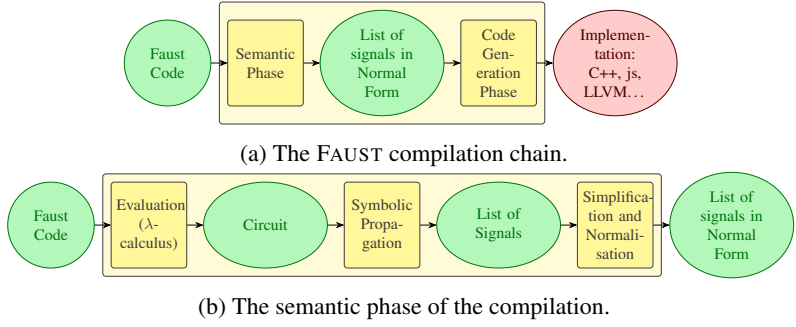

The \F compiler is composed of several steps, which are illustrated in Figure~\ref{fig:faustchain}:
starting from the DSP source code, the Semantic Phase produces signals as conceptually infinite streams of samples or control values.
Those signals are then compiled to imperative code (C/C++, LLVM IR, WebAssembly, etc.) in the Code Generation Phase.

The Semantic Phase itself, zoomed on in Figure~\ref{fig:semanticphase}, is composed of several steps:
the initial DSP code using the Block Diagram Algebra is translated into a flat circuit in normal form in the lambda-calculus evaluation step. 
The list of output signals is produced by the Symbolic Propagation step.
Each output signal is then simplified and a set of optimisations are done
(normal form computation and simplification, delay line sharing, typing, etc.)
to finally produce a list of output signals in normal form.
During this phase, a number of properties over the signals can be computed in a recursive fashion.
The Code Generation Phase translates the signals in an intermediate representation named FIR (\F Imperative Representation)
which is then converted to the final target language (C/C++, LLVM IR, WebAssembly,etc.) with a set of backends.

\begin{figure*}[t]
  \centering

  \begin{subfigure}[b]{0.5\textwidth}
      \centering
      \begin{lstlisting}
process = +(1/64) ~ %(1)
      : *(2*ma.PI) : sin;
      \end{lstlisting}
    \caption{\F source code}
  \end{subfigure}
  \begin{subfigure}[b]{0.4\textwidth}
    \includegraphics[width=\textwidth]{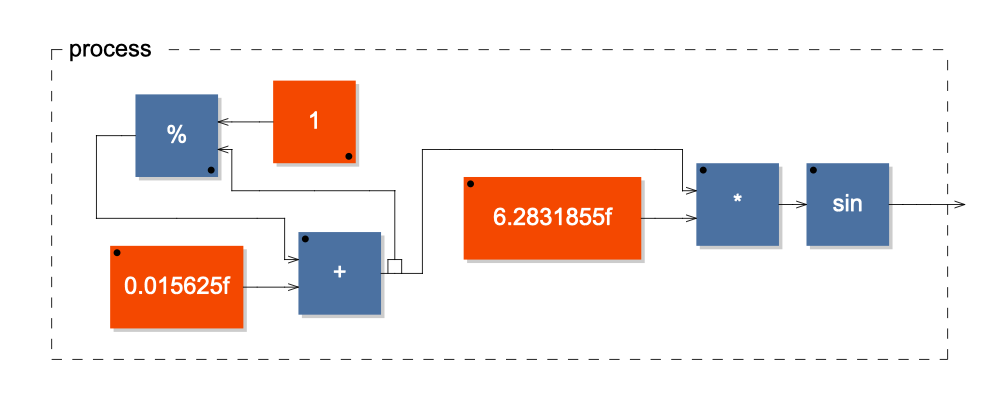}
    \caption{\F block diagram}
  \end{subfigure}

  \vfill
  
  \begin{subfigure}[b]{\textwidth}
    \centering
    \includegraphics[width=\textwidth]{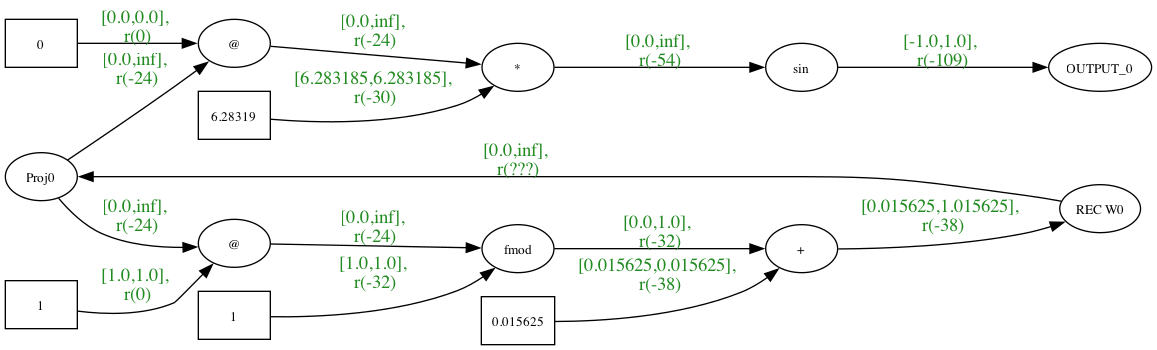} 
    \caption{Signal graph indicating inferred ranges and precisions (denoted \lstinline{r}, for {\em resolution}) of signals.}
  \end{subfigure}
    
    \hfill

    \begin{subfigure}{\textwidth}
  \begin{subfigure}[b]{0.6\textwidth}
    \begin{lstlisting}
fRec0[0] = sfx_t(31,-24)(sfx_t(1,-38)(
          sfx_t(0,-32)(fmodfx(fRec0[1], sfx_t(0,-32)(1.0)))
        + sfx_t(-6,-38)(0.015625)));
output0[i0] = FAUSTFLOAT(sfx_t(0,-109)(
          sinfx(sfx_t(31,-54)(sfx_t(3,-30)(6.2831855)
        * fRec0[0]))));
fRec0[1] = fRec0[0];
\end{lstlisting}        
\caption{Generated fixed-point code, annotated with casts to fixed-point datatypes.}
  \end{subfigure}
    \begin{subfigure}[b]{0.3\textwidth}
      \includegraphics[width=0.9\textwidth]{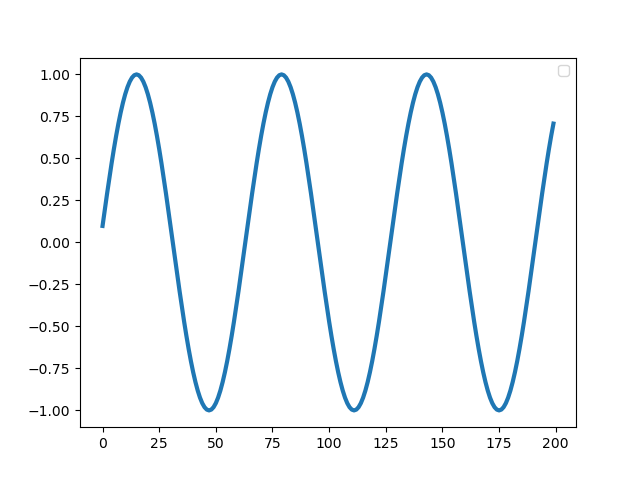}
      \caption{Output sinewave}
    \end{subfigure}
  \end{subfigure}

  \caption{Compilation of a phasor-based sine, from the \F code to the audio output.
    Range of signals and prescribed precisions are determined during the Semantic Phase of the compilation,
    converting the block diagram to a signal graph.
    These signal properties are then used to determine fixed-point formats,
    which are indicated in the generated code as casts to fixed-point datatypes.}
  \label{fig:range_prop}
\end{figure*}

Figure~\ref{fig:range_prop} illustrates the propagation of a signal range through a program.
For example, one can observe that the output range of \lstinline{fmod} is $[0;1[$
because its second argument is $1$, and that the output range of \lstinline{sin} is $[-1;1]$
because its input range spans the $2\pi$ period of \lstinline{sin}.

The ranges and precisions computed on the signal level are converted to fixed-point formats,
represented in the code by the \lstinline{sfx_t} macro,
which serves to translate the $(m, \ell)$ format used in this work
to the $(w, m)$ format used by the~\lstinline{ap_fixed} library~\cite{apfixed},
which we use for our implementation.
Values in the code are constrained to their prescribed formats through casts.

\subsection{The intervals library}

Intervals of signals are a property computed during the symbolic propagation phase.
As presented in Section~\ref{sec:range}, these intervals describe the range of values that a signal can potentially take.
Intervals are present in the compiler as annotations on the signals.

An interval is defined by three fields: its lowest value, its highest value,
and the precision of the interval, given as the position of the LSB.

Interval inference is done from inputs to outputs.
The intervals of source signals (leaves of the signal graph) can be readily inferred:
audio inputs take values in the $[-1;1]$ interval;
a numerical constant with value $K \in \RR$ takes values in the singleton $[K;K]$;
a slider provided with minimum and maximum values $x_{min}$ and $x_{max}$ will give values in the $[x_{min}; x_{max}]$ interval.

When a signal is fed into a real function $f$,
the interval of the output signal is the smallest interval (that is, continuous subset of $\RR$) containing the image of the interval under $f$.
For example, let us consider the inverse function $\mathrm{Inv}: x \mapsto \frac{1}{x}$.
The image of $[-1;1]$ under $\mathrm{Inv}$ is $\mathrm{Inv}([-1; 1]) = [-\infty; -1] \cup [1; +\infty]$,
which is not an inverval: there is a discontinuity in $]-1;1[$.
The smallest interval containing $\mathrm{Inv}([-1;1])$ is $[-\infty; +\infty]$.

The interval library was primarily introduced to bound the size of delay lines
in order to generate code with a properly-sized buffer.
It is also used to detect range violations,
such as division by $0$ or negative delay.
In the context of fixed-point format determination,
it is used to determine the position $m$ of the MSB:
for a signal with range $[\xlo;\xhi]$, $m = \ceil{\log_2(max(\abs{\xlo}, \abs{\xhi}))}$.
As will be developed in next section, the range of a signal is also one of the parameters
used to determine the LSB position to assign to it.

\subsection{Precision inference}

While the bounds of the interval can be readily inferred using interval arithmetic,
one has to establish a criterion linking input and output precision in order to infer one from the other.

\subsubsection{Pseudo-injectivity}

We introduce a criterion that ensures that all output values have enough bits
to represent all the information contained in input values,
and formalise it under the notion of pseudo-injectivity.

Let us first introduce a few notations:
\begin{itemize}
\item $[\xlo; \xhi]_\ell = [\xlo; \xhi] \cap 2^\ell \ZZ$ denotes all the values
  in the $[\xlo; \xhi]$ interval representable with precision $\ell$.
\item $u = 2^\ell$ denotes the gap between two consecutive numbers representable with precision $\ell$.
  Notation modifiers will be matched: for example, $u' = 2^{\ell'}$.
\item $\lfloor x \rfloor_{\ell} = u\cdot\floor{\frac{x}{u}}$ denotes the rounding down of $x$ to the closest number with precision $\ell$.
\end{itemize}

\begin{defi}
  The input LSB $\ell$ and the output LSB $\ell'$ of a function $f : [\xlo; \xhi] \to \RR$ are said
  to respect the pseudo-injectivity condition if
  \begin{equation}
    \begin{split}
      \forall x_1, x_2 \in [\xlo;\xhi], \lfloor x_1 \rfloor_{\ell} \neq \lfloor x_2 \rfloor_{\ell}
      \\ \Rightarrow \lfloor f(x_1) \rfloor_{\ell'} \neq \lfloor f(x_2) \rfloor_{\ell'} \textrm{ or } f(x_1) = f(x_2)
    \end{split}
  \end{equation}
\end{defi}

\begin{figure}
  \centerline{\framebox{
      \includegraphics[width=\columnwidth]{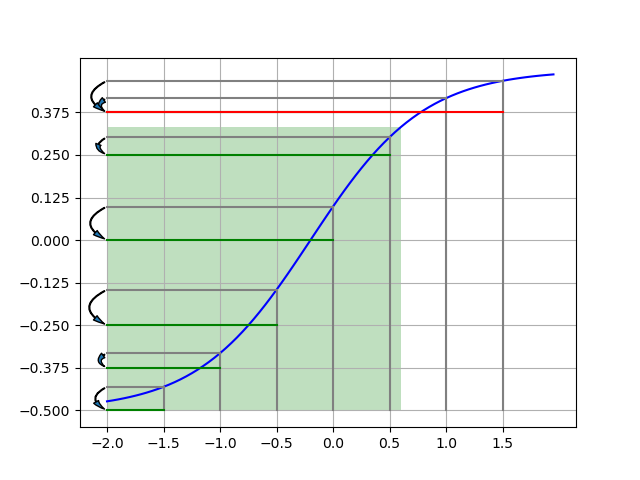} 
    }}
  \caption{Graph of a function defined over $[-2; 2[$ with its input interval discretised with a $0.5$ step
    and its output interval discretised with a $0.125$ step.
    The real images of the discrete inputs are projected on the output axis,
    and arrows indicate which discrete image they are rounded to.
    Rounded values of images are indicated by a green line if only one image is rounded to a given value,
    and a red line if more than two images are rounded to the same value.}
  \label{fig:pseudoinj}
\end{figure}

Intuitively, our goal that, if two real input values $x_1$ and $x_2$ are discretised into two different fixed-point values
($\lfloor x_1 \rfloor_{\ell} \neq \lfloor x_2 \rfloor_{\ell}$),
then their images $f(x_1)$ and $f(x_2)$ through function $f$ should be discretised into distinct fixed-point values as well
($\lfloor f(x_1) \rfloor_{\ell'} \neq \lfloor f(x_2) \rfloor_{\ell'}$),
unless these images were mathematically identical to begin with ($f(x_1) = f(x_2)$).

For example, Figure~\ref{fig:pseudoinj} illustrates the discretisation of $f: x \mapsto \frac{1}{2}\cdot \tanh(x + \frac{1}{5})$
with input precision $0.5$ and output precision $0.125$.
This discretisation is enough to correctly represent the images of values in the $[-2.0; 0.5]$ interval,
but outside of it,  $f(1.0) \approx 0.4168$ and $f(1.5) \approx 0.4677$ are collapsed into the single value $0.375$.

Ideally, to follow an economy principle, the output precision should be the lowest one respecting pseudo-injectivity.

We established a closed formula to compute the optimal output precision $\ell'$
from the function $f$, the input interval $[\xlo; \xhi]$ and the input precision $\ell$:
\begin{equation}\label{eq:cf_pseudoinj}
  \ell' = \Min_{\substack{x \in [\xlo; \xhi]_{\ell},\\ f(x) \neq f(x+u)}} \floor{\log_2(\abs{f(x + u) - f(x)})}
\end{equation}

This formula looks for the smallest, non-null difference existing between two consecutive images of discrete arguments,
and yields the bit position corresponding to the amplitude of this difference.
If the difference between the values is representable,
then their finite-precision representations should be distinct.

In practice, the computation of output precision is done by approximating the error through a Taylor expansion:
$f(x+u) - f(x) \approx u f'(x)$,
so $\floor{\log_2(\abs{f(x + u) - f(x)})} \approx \ell + \log_2(\abs{f'(x)})$.
The implemented computation of $\ell'$ uses the point of lowest value of $\abs{f'}$,
which does not exactly correspond to the minimum of $\abs{f(x+u) - f(x)}$,
but is a good enough approximation in practice.

Pseudo-injectivity is preserved by function composition,
which makes it a relevant property to propagate through the signal graph.

\begin{prop}
  If $f: [\xlo;\xhi]_{\ell} \rightarrow [\ylo;\yhi]_{\ell'}$ and $g: [\ylo;\yhi]_{\ell'} \rightarrow [\zlo;\zhi]_{\ell''}$ are pseudo-injective,
  then $g\circ f:[\xlo;\xhi]_{\ell} \rightarrow [\zlo;\zhi]_{\ell''}$ is pseudo-injective as well.
\end{prop}

Let us now evaluate the coherence of this method with the error analysis presented in Section~\ref{sec:accuracy}.
If a real input variable $x$ presents an error $\epsilon$,
then the output error is $\abs{f(x + \epsilon) - f(x)}$.
If we only take into consideration the input error due to rounding, we have $\epsilon \sim u$,
which is constant over the input range.
This error becomes $uf'(x)$ in the output.
The magnitude of the output error is different on different points of the output interval,
which compels us to decide which one should be taken into account to determine output precision.
The first impulse is to seek to eliminate all error by aligning $\ell'$
with the point of highest error ({\em i.e.} the maximum of $\abs{f'}$).
However, doing so also eliminates information-containing bits in zones of the interval where the error is lower.
In addition, preliminary experimentation suggests that this approach yiels poor-quality results.
The approach we settled for instead seeks the point of lowest error,
which allows to guarantee that all the dismissed bits contain error,
and that no valid information is eliminated by rounding.

\subsubsection{Precision computation}

The pseudo-injectivity property is then propagated through the signal graph at the same time as the intervals,
from inputs to outputs.
Audio inputs are assumed to have the precision set by ADCs (generally $(8, -24)$ format).
Constants are given a fixed width of $32$.
The precision of sliders is inferred from the set step: a slider with step $\alpha$ implies $\ell = \floor{\log_2(\alpha)}$.
Precision is then propagated from the inputs to the output of each function implicated in the signal graph using Equation~\ref{eq:cf_pseudoinj}.

In most cases, recursions do not have an invariant when it comes to precision:
often, applying Equation~\ref{eq:cf_pseudoinj} to the contents of one recursion iteration
will strictly increase or decrease the prescribed precision.
If taken at face value, this will lead to infinite precisions, which are impracticable.
Hence, we chose to uniformly assign a precision of $-24$ to recursive constructs.
This is bound to change in the future as we uncover finer precision inference methods for recursive programs.

This first propagation step uses a structure that is already present in the compiler, 
allowing to propagate from inputs to outputs.
However, it also tends to result in large values of $\ell$ for formats throughout the graph.
Future work will be needed to address this issue.

\section{Results}\label{sec:results}

Let us now present and discuss the results of this algorithm on two examples:
the sinewave presented in Figure~\ref{fig:range_prop} and the Karplus-Strong string synthesis algorithm.

\begin{table}
\begin{center}
\begin{tabular}{|l|c|}
\hline
Program & Signal-to-noise ratio \\
\hline
Sine with increment $\frac{1}{64}$  & 32 \\
\hline
Sine with increment $0.01$  & 25 \\
\hline
Karplus-Strong  & 33 \\ 
\hline
\end{tabular}
\end{center}
\caption{Signal-to-noise ratios quantifying discrepancies between fixed-point and floating-point versions of test programs.}
\label{tab:snr}
\end{table}

\begin{figure}
  \centering\framebox{
  \begin{subfigure}[b]{0.45\columnwidth}
    \includegraphics[width=\textwidth]{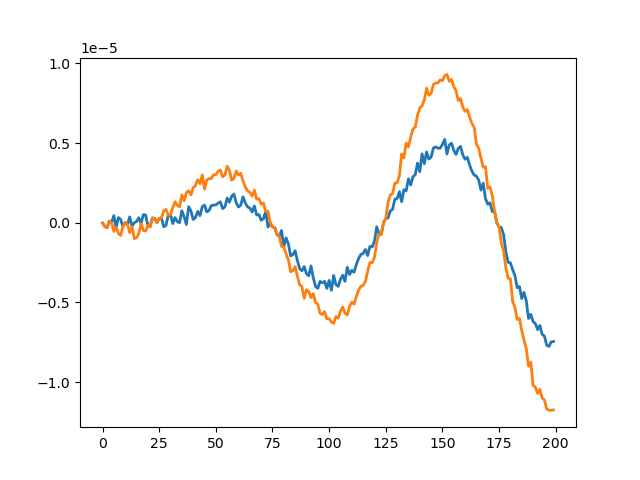}
    \caption{Errors committed on a sinewave with increment 0.01.
      Due to the increment not being exactly representable in binary,
      the difference with the ground truth increases indefinitely.}
  \end{subfigure}}
  \framebox{\begin{subfigure}[b]{0.45\columnwidth}
    \includegraphics[width=\textwidth]{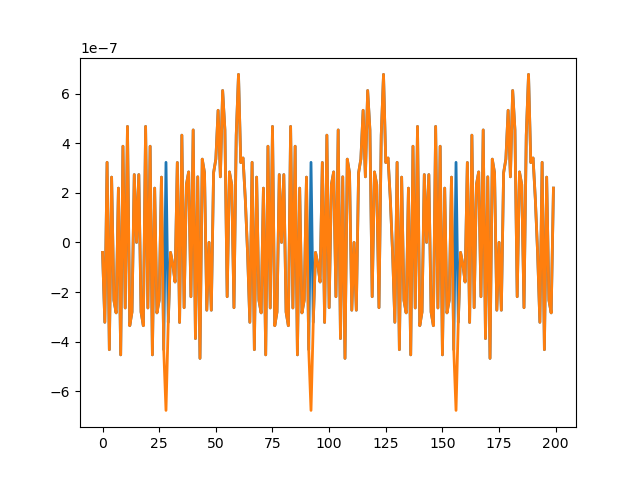}
    \caption{Errors committed on a sinewave with increment $\frac{1}{64}$.
    The difference with the ground truth exists, but it is confined to an amplitude of around $6\times 10^{-7}$.}
  \end{subfigure}}
  \caption{Errors committed by floating-point (in blue) and fixed-point (in orange) implementations
    of the phasor sinewave presented in Figure~\ref{fig:range_prop}. The ground-truth for the error is computed with numpy.}
  \label{fig:sineerror}
\end{figure}

Table~\ref{tab:snr} presents signal-to-noise ratios achieved by the test programs discussed in the next sections.
We denote by $s_i$ the samples computed by the floating-point version of the program and $\hat{s_i}$ their fixed-point counterpart.
We define here the signal-to-noise ratio as $\log_{10}(\frac{S}{N})$,
where $S = \Sum_{i=1}^{200} s_i^2$ is the power of the signal, 
and $N = \Sum_{i=1}^{200} (s_i - \hat{s_i})^2$ is the power of the noise over a window of 200 samples.
Considering the floating-point implementation of an algorithm as a ground truth for our signal is technically false
(Figure~\ref{fig:sineerror} shows that floating-point also incurs errors),
but it is useful to quantify how much the fixed-point version of a program deviates from its floating-point counterpart,
assumed to be soundly implemented.

\subsection{Sinewave}

The \F program presented in Figure~\ref{fig:range_prop} outputs a sine-wave. The first part of the program, the recursive construct \lstinline{+(1/64) ~ 

The choice of the increment (here $\frac{1}{64}$) and the sampling frequency of the audio output
both factor in the end frequency of the sinewave.
Here, the chosen value $\frac{1}{64} = 2^{-6} = 0.015625$ could be represented over one single bit, with format $(-6, -6)$.
However, this precision choice gets passed on downstream in the signal graph,
and ultimately influences the precision used for the audio output.
If we instead use the numerically close value $0.02$, which is not finitely representable in binary,
the act of constraining it to format $(-6, -6)$ rounds it to $2^{-6}$,
rendering the program undistinguishable from the other version.
On the contrary, using a more generous format such as $(-6, -38)$
allows for a more accurate representation of the constant $0.02$,
but leads to a different precision for the audio output than the first version of the program.
This seems to indicate that constants' precisions should not be optimised before they are propagated.

Overall, the precisions inferred for this program are very large,
resulting in particular in the audio output being assigned a precision of $\ell = -109$.
This is a much larger bitwidth than the standard $16$ or $24$ bits audio formats.
This hints that further steps to compute lower precisions, compatible with a $16$ or $24$ bits audio format,
and that still respect pseudo-injectivity, will be needed.

\subsection{Karplus-Strong}

Our second example is a simplified implementation of the Karplus-Strong algorithm,
synthesising a string sound. The code is presented in Figure~\ref{fig:karplus-faust}.
\begin{figure}\centering
  
  \begin{lstlisting}
    process = 1 - 1' : + ~ filter
    with
    {
      filter = @(del): avg;
      avg(x) = (x+x')/2;
      del = 50;
    };
  \end{lstlisting}
  \caption{\F implementation of the Karplus-Strong algorithm.}
  \label{fig:karplus-faust}
\end{figure}

This algorithm works by feeding an excitation signal into a feedback loop.
Here, we use an impulsion \lstinline{1-1'} as an excitation,
which is not as efficient as the usual white noise, but has the advantage of simplicity
and does not unduly complexify the analysis.
The feedback loop \lstinline{filter} is made of a low-pass filter delayed by 50 samples.
To keep the program simple and its analysis tractable,
we use the very simple low-pass filter \lstinline{avg} which averages the values of two consecutive samples.

The generated code is given in Figure~\ref{fig:karplus-gen}.

\begin{figure}\centering
    
    \begin{lstlisting}
iVec0[0] = 1;
fRec0[IOTA0 & 63] =
 sfx_t(31,-24)(sfx_t(31,-57)(
   sfx_t(0,-24)(sfx_t(0,-24)(1 - iVec0[1]))
 + sfx_t(31,-57)(sfx_t(-1,-33)(0.5)
   * sfx_t(31,-24)(
      sfx_t(31,-24)(fRec0[(IOTA0 - 51) & 63])
    + sfx_t(31,-24)(fRec0[(IOTA0 - 52) & 63])))
  ));
output0[i0] = FAUSTFLOAT(
  sfx_t(31,-24)(fRec0[IOTA0 & 63]));
iVec0[1] = iVec0[0];
IOTA0 = IOTA0 + 1;
\end{lstlisting}
\caption{Generated fixed-point code for the Karplus-Strong algorithm.}
\label{fig:karplus-gen}
\end{figure}

The constant $0.5$, present as \lstinline{1/2} in the \lstinline{avg} low-pass filer,
is represented over $32$~bits with the $(-1, -33)$ format.
The delayed feedback line is represented with $\ell = -24$ bits of precision.
The sum of two consecutive samples is represented with the same precision,
and, when multiplied by $0.5$, the result is represented with $-57 = -24 + (-33)$ bits of precision.
The final audio output is represented with $-57$ bits of precision,
which is more precise than the usual audio formats, 
and indicates that there is still room for optimisation.

\section{Conclusion}

In this paper, we laid out the basics of floating-point and fixed-point formats,
and the challenges posed by their respective use in the field of digital audio.

We presented a method for the inference of adapted fixed-point formats in \F programs,
introducing the notion of pseudo-injectivity and presenting its implementation
in the \F compiler.

The preliminary tests of our method give promising results in terms of preservation of audio quality,
but tends to infer formats wider than strictly needed.
This seems to indicate that further improvements are possible.

\subsection{Future work}

\subsubsection{Backwards propagation}

As remarked in Section~\ref{sec:results}, the formats inferred for audio outputs tend to be much wider than usual audio formats.
After having propagated precisions from program inputs to outputs, a second pass of propagation could be carried out, this time from outputs to inputs.
It would begin by setting standard audio formats for the outputs and working its way up by determining precisions compatible with pseudo-injectivity.

We have not implemented such backward propagation yet. However, preliminary results computed by hand,
seem to indicate that this would result in lower bitwidth throughout the program.

\subsubsection{Errors as intervals}

As explained in Section~\ref{sec:accuracy}, errors propagating in programs can be modelled as intervals.
While the precisions have been initially implemented in \F programs as integers,
it would be straightforward to swap them for intervals, allowing for a finer representation of precision.

\subsubsection{Targeted optimisation}

Besides this general precision propagating principle, other, more targeted, optimisations can be carried out.

\F programmers typically write sliders using powers of~$10$ for stepsizes
($0.1$, $0.01$, etc).
These values are convenient to write in a program,
but are not finitely representable in binary.
In most cases, such stepsizes could be set to a power of~$2$ with the same order of magnitude
(for example, replacing $0.1$ with $0.125 = 2^{-3}$),
without significantly perturbating the semantics of the program.

Another avenue for optimisation would be the detection of structures of interest in \F programs.
For example, the \F compiler could detect expressions of the form $\sin(\pi\cdot x)$,
and rewrite them to use the function $\textrm{sinPi} : x \mapsto \sin(\pi \cdot x)$,
a standard primitive especially implemented to correctly deal with the irrationality of $\pi$.

Work is currently in progress to integrate FIR and IIR filters structures as \F primitives
and detect them in the normalisation phase of the compiler.
As discussed in Section~\ref{sec:accuracy}, LTI filters are well-studied from the point of view of computer arithmetics.
Detecting IIR filters as a primitive would allow to exploit these results.

\subsubsection{Probabilistic precision}

We have also studied the possibility of relaxing the notion of pseudo-injectivity,
to allow for a portion of discrete images to be undistinguishable.
This would allow for lower bitwidths in programs, with a tunable parameter
controlling the proportion of images one is prepared to sacrifice.

\bibliographystyle{plain}
\bibliography{biblio,software}

\end{document}